\documentclass[12pt,a4paper,dvips]{article}

\usepackage{a4p}
\usepackage{cite,mcite}
\usepackage{graphicx}
\usepackage{physics}
\usepackage{amssymb,amsbsy,hhline}
\usepackage{l3_title,ifthen,Lep}

\journalname{Phys. Lett. B}
\date{26 January 2001}

\preprint{2001-007}

\Lep{1}

\newlength{\capindent}
\newlength{\capwidth}
\newlength{\figwidth}
\newcommand{\icaption}[2][!*!,!]{\hspace*{\capindent}%
  \begin{minipage}{\capwidth}
    \ifthenelse{\equal{#1}{!*!,!}}%
      {\caption{#2}}%
      {\caption[#1]{#2}}
  \end{minipage}}

\def\Vud{\ensuremath{\vert V_{\mathrm{ud}} \vert}}
\def\Vus{\ensuremath{\vert V_{\mathrm{us}} \vert}}
\def\als{\ensuremath{\alpha_s}}
\def\alsopi{\left(\frac{\als}{\pi}\right)}
\def\antibar#1{#1\bar{#1}}
\def\brel {\ensuremath{{\cal{B}}(\tenn)}}
\def\brhad{\ensuremath{{\cal{B}}(\tau\rightarrow h \nu_{\tau})}}
\def\brlep{\ensuremath{{\cal{B}}(\tau\rightarrow\ell\bar{\nu}_{\ell}\nu_{\tau})}}
\def\brmu {\ensuremath{{\cal{B}}(\tmnn)}}
\def\dnpqcd{\ensuremath{\delta_{NP}}}
\def\gf{\ensuremath{G_F}}
\def\gwe{\ensuremath{g_{\e}}}
\def\gwl{\ensuremath{g_{\ell}}}
\def\gwm{\ensuremath{g_{\mu}}}
\def\gwt{\ensuremath{g_{\tau}}}
\def\qqbar{\antibar{\mathrm{q}}}
\def\ffbar{\antibar{\mathrm{f}}}
\def\rtau{\ensuremath{R_{\tau}}}
\def\sew{\ensuremath{S_{EW}}}
\def\tenn{\ensuremath{\tau\rightarrow\e\bar{\nu}_{\e}\nu_{\tau}}}
\def\tlnn{\ensuremath{\tau\rightarrow\ell\bar{\nu}_{\ell}\nu_{\tau}}}
\def\tmnn{\ensuremath{\tau\rightarrow\mu\bar{\nu}_{\mu}\nu_{\tau}}}
\def\ttau{\ensuremath{\tau_{\tau}}}
\def\eeee{\ensuremath{\ee\!\rightarrow \ee(\gamma)}}
\def\eemm{\ensuremath{\ee\!\rightarrow \mu^+\mu^-(\gamma)}}
\def\eeqq{\ensuremath{\ee\!\rightarrow \qqbar}}
\def\eett{\ensuremath{\ee\!\rightarrow \tau^+\tau^-(\gamma)}}
\def\ggee{\ensuremath{\ee\!\rightarrow \ee\ee}}
\def\ggmm{\ensuremath{\ee\!\rightarrow \ee\mu^+\mu^-}}
\def\ggff{\ensuremath{\ee\!\rightarrow \ee\ffbar}}
\def\ggqq{\ensuremath{\ee\!\rightarrow \ee\qqbar}}

\def\STAT{\ensuremath{\mathrm{stat.}}}
\def\SYST{\ensuremath{\mathrm{syst.}}}

\graphicspath{{./}}

\begin{document}
\begin{titlepage}

\title{Measurement of the Tau Branching Fractions \\ into Leptons}

\author{The L3 Collaboration}

\begin{abstract}

Using  data  collected  with  the  L3  detector  near  the Z  resonance,
corresponding  to  an  integrated   luminosity  of  150  pb$^{-1}$,  the
branching  fractions  of the tau  lepton  into  electron  and  muon  are
measured to be
\begin{eqnarray*}
 & \brel & = (17.806 \pm 0.104 (\STAT) \pm 0.076 (\SYST) ) \%, \\
 & \brmu & = (17.342 \pm 0.110 (\STAT) \pm 0.067 (\SYST) ) \%.
\end{eqnarray*}
From these results the ratio of the charged current  coupling  constants
of the muon and the electron is determined to be $\gwm/\gwe = 1.0007 \pm
0.0051$.  Assuming  electron-muon  universality,  the Fermi  constant is
measured  in tau  lepton  decays as $\gf = (1.1616  \pm  0.0058)  \times
10^{-5}  \GeV^{-2}$.  Furthermore,  the coupling  constant of the strong
interaction  at the tau  mass  scale is  obtained  as  $\als(\Mtau^2)  =
0.322\,\pm\,0.009\,\mbox{(exp.)}\,\pm\,0.015\,\mbox{(theory)}$.

\end{abstract}

\submitted

\end{titlepage}

\section*{Introduction} 

In        the        Standard         Model        of        electroweak
interactions~\cite{GWS1},  the  couplings  of the leptons to
the gauge bosons are assumed to be independent of the lepton generation.
Measurements  of the leptonic  branching  fractions  $\brlep$ and of the
lifetime  $\ttau$ of the tau lepton provide a test of this  universality
hypothesis  for the  charged  current.  The  leptonic  width  of the tau
lepton~\cite{Marciano88},
\begin{eqnarray}
\Gamma(\tlnn) & \equiv & \frac{\brlep}{\ttau} \\
              & = & \frac{\gwt^2\gwl^2}{\MW^4}\frac{\Mtau^5}{96 (4\pi)^3}
                    (1+\epsilon_P)\,(1+\epsilon_\mathrm{QED})\,(1+\epsilon_{q^2}),
\end{eqnarray}

\noindent where $\ell =$ e, $\mu$, depends on the coupling  constants of
the tau lepton and the lepton  $\ell$ to the W boson, $\gwt$ and $\gwl$,
respectively.  Here  $\Mtau$  and $\MW$ are the masses of the tau lepton
and the W boson.  The quantities  $\epsilon_P$,  $\epsilon_\mathrm{QED}$
and  $\epsilon_{q^2}$ are small corrections  resulting from phase-space,
QED         corrections~\cite{Kino_Sir}         and         the        W
propagator~\cite{Ber_Sir,Riemann}, respectively.

The  comparison of the tau branching  fractions  into  electron and muon
gives a direct  measurement  of the  ratio  $\gwm/\gwe$.  Moreover,  tau
decays into hadrons are sensitive to the strong coupling constant $\als$
at the tau mass  scale.  The  ratio of the  hadronic  to the  electronic
width,  $\rtau$,  can be  expressed in terms of the  leptonic  branching
fractions:

\begin{equation}
   \label{eq:rtau}
   \rtau = \frac{\brhad}{\brel} = \frac{1-\brel-\brmu}{\brel},
\end{equation}

\noindent  where  $\brhad$ is the branching  fraction of the $\tau$ into
hadrons.      $\rtau$      is      calculated      in       perturbative
QCD~\cite{Gorishny91,Kataev95} as:
\begin{eqnarray}
  \nonumber
  \rtau & = & 3 \, \left(\Vud^2 + \Vus^2\right) \, \sew \\
        & \times & \left( 1 \, + \, \frac{\als}{\pi} \, + \,
           5.2023 \alsopi^2 \, + \, 
           26.366 \alsopi^3 \, + \,
          (78+d_3)\alsopi^4 \, + \, \dnpqcd \right),
  \label{eq:als}
\end{eqnarray}

\noindent where  $V_{\mathrm{ud}}$ and $V_{\mathrm{us}}$ are elements of
the       Cabibbo-Kobayashi-Maskawa       (CKM)       quark       mixing
matrix~\cite{ckm1}.  The  quantities   $\sew$~\cite{Sirlin82}  and
$\dnpqcd$~\cite{neubert96}  describe short range  electroweak  radiative
corrections and non-perturbative  QCD contributions,  respectively.  The
quantity  $d_3$ is  estimated  to be  27.5~\cite{Kataev95},  to which we
assign a 100\% uncertainty.

This paper presents a measurement  of the tau branching  fractions  into
electron and muon with the L3  detector~\cite{l3_det}  at LEP using data
taken  near  the  $\Zo$  pole.  The  results  supersede  our  previously
published   ones~\cite{oldl3}.  Results  from  other   experiments   are
reported elsewhere~\cite{otherbr}.

\section*{Data Sample and Monte Carlo Simulation}

The data  were  collected  from  1991 to  1995\footnote{In  this  letter
figures  and  tables  often  refer just to the 1994  data,  which is the
largest  sample.}  at  centre-of-mass  energies  around  the Z mass  and
correspond to an integrated luminosity of about 150 pb$^{-1}$.

For  efficiency  and  background   estimates,  Monte  Carlo  events  are
generated  using the  programs  KORALZ~\cite{koralz1}  for  $\eemm$  and
$\eett$,       BHAGENE~\cite{bhagene1}       for      $\eeee$,
DIAG36~\cite{diag36} for $\ggff$, where $\ffbar$ is $\ee$, $\mu^+\mu^-$,
$\tau^+\tau^-$    or   $\qqbar$,   and    JETSET~\cite{my_jetset}    for
$\ee\ra\qqbar(\gamma)$.  The Monte  Carlo  events are  passed  through a
full detector  simulation,  based on the GEANT  program~\cite{my_geant},
which  takes  into  account  the  effects  of  energy   loss,   multiple
scattering, showering and time dependent detector inefficiencies.  These
events are  reconstructed  with the same program used for the data.  The
number of Monte Carlo  events in each process is about ten times  larger
than the corresponding data sample.

\section*{Measurement Technique}
  
To  measure  the tau  leptonic  branching  fractions,  first a sample of
$\eett$  events is selected  with some  remnant  background  mainly from
other leptonic Z decays.  From this sample the branching fraction of the
tau into a lepton $\ell$ is then obtained as:
\begin{eqnarray}
   \brlep = \frac{N_\ell(1-f_\ell^{non \;\tau}-f_{h \to l}^\tau)}
            {N_\tau(1-f_\tau^{non \;\tau})}\,
            \frac{1}{\varepsilon^{ID}_\ell}\,
            \displaystyle {\frac{\varepsilon^{sel}_\tau}{\varepsilon^{sel}_\ell}},
\label{brform}
\end{eqnarray}
where    $N_\tau$    is   the   number   of    selected    tau   decays,
$\varepsilon^{sel}_\tau$    is   the   $\eett$   selection   efficiency,
$f_\tau^{non   \;\tau}$  is  the   background  of  other  final  states,
$\varepsilon^{sel}_\ell$ is the selection efficiency for tau decays into
the  lepton  $\ell$,  $N_\ell$  the  number of  identified  leptons  and
$\varepsilon^{ID}_\ell$  the  identification  efficiency  of the  lepton
$\ell$.  The quantities $f_{h \to l}^\tau$ and $f_\ell^{non \;\tau}$ are
the background  contaminations  from other $\tau$ decays and  non-$\tau$
final states in the selected leptonic decays, respectively.

Systematic     uncertainties    from    the    selection    efficiencies
$\varepsilon^{sel}_\tau$  and  $\varepsilon^{sel}_\ell$  cancel  if  the
ratio  $\varepsilon^{sel}_\tau /  \varepsilon^{sel}_\ell$  equals unity,
{\it i.e.}  the  selection of $\eett$  does not  introduce a bias to the
fractions of leptonic  tau decays.  This bias is avoided by  subdividing
each event into two hemispheres by a plane  perpendicular  to the thrust
axis.  Then, the  selection  of $\eett$  events is  performed  using the
information   from  just  one   hemisphere,   called  the  {\it   tagged
hemisphere}.  This  information  is not  correlated  with  the  opposite
hemisphere, denoted as the {\it analysis hemisphere}, that is subject to
electron and muon  identification.  The $\tau$ decays  identified in the
analysis  hemisphere  constitute  a bias free  sample for the  branching
fraction measurement.

\section*{Selection of \mathversion{bold}$\eett$\mathversion{normal} Events}

Events of the process $\eett$ are  characterised by two low multiplicity
jets which are almost  back-to-back.  To ensure good track  measurements
in the central tracker, the fiducial volume is defined by $|\cos \theta|
<$ 0.72,  where  $\theta$  is the polar  angle of the  thrust  axis with
respect  to the  electron  beam  direction.  The  requirements  for  the
preselection of leptonic Z decays are:
\begin{itemize}
\item{the number of tracks must be less than 10,}
\item{the number of tracks in each hemisphere must be less than 7,}
\item{each  hemisphere  must have at least one track  with a  transverse
      momentum greater than 1~\GeV,}
\item{the distance of closest approach of at least one track to the beam
      position in the plane perpendicular to the beam axis, must be less
      than 5 mm.}
\end{itemize}

\begin{figure}
    \includegraphics[width=\textwidth]{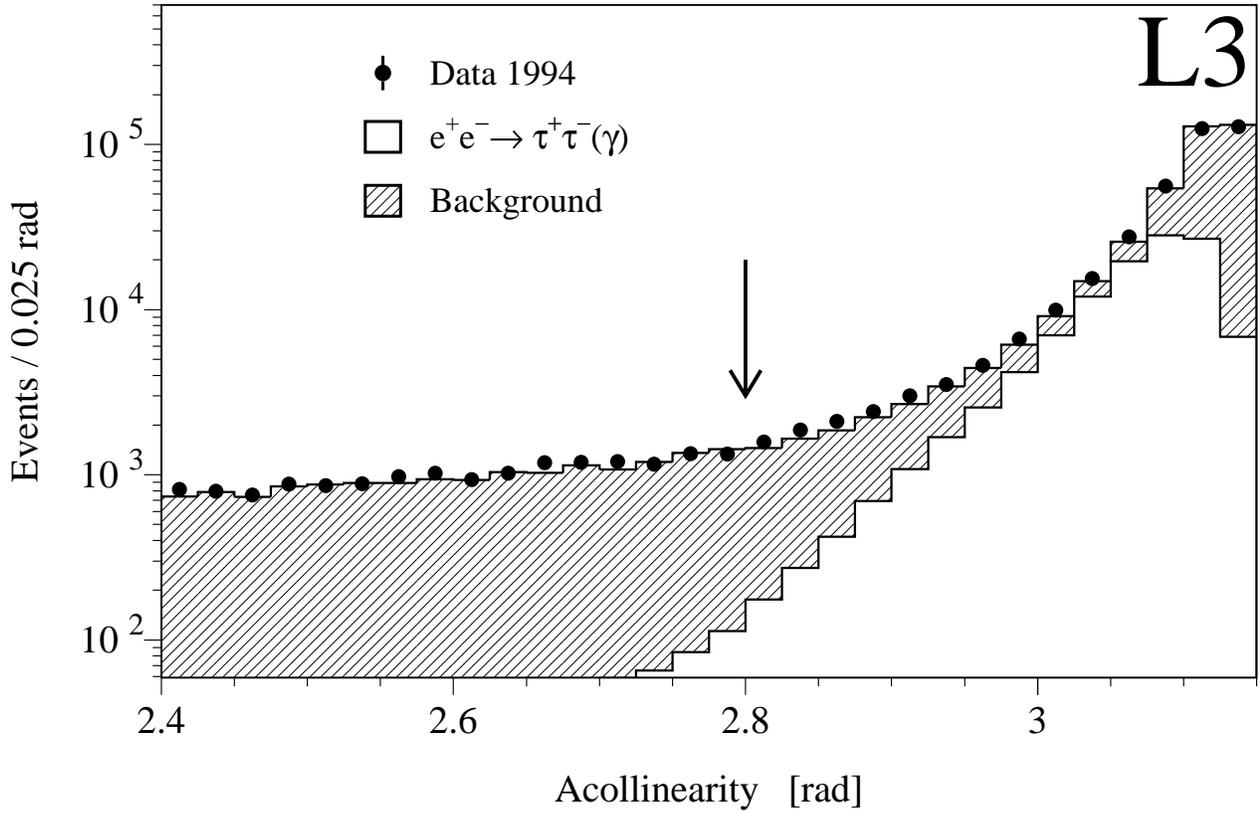}
    \caption[]{\label{fig:acol}  Distribution of the acollinearity after
                the  preselection.  The arrow  indicates the position of
                the cut applied to select $\eett$ events.}
\end{figure}

The only  additional  requirement  on the  event  as a whole is that the
acollinearity  angle between the leading  tracks of the two  hemispheres
must be larger than 2.8~rad.  The background suppression due to this cut
is  illustrated  in  Figure~\ref{fig:acol}   for  1994  data  after  the
preselection.

The criteria to select $\eett$ events in the tagged hemisphere are:
\begin{itemize}  
\item{there must be one, two or three tracks in the central tracker,}

\item{the angle between each track and the thrust axis must be less than
      0.45 rad,}

\item{the  energy of the most energetic  cluster in the  electromagnetic
      calorimeter, BGO, must be less than 75\% of the beam energy,}

\item{the momentum of a track reconstructed in the muon chambers must be
      less than 65\% of the beam energy.}
\end{itemize}
                                                              
Background  events that fall into less efficient regions of the detector
can fake $\eett$ events.  They are identified by the following  criteria
in the tagged hemisphere:
\begin{itemize}  

\item{\eeee}: there is one track  pointing to the carbon  fibre  support
              between crystals of the electromagnetic  calorimeter, with
              either its momentum or the  corresponding  energy measured
              in the BGO larger than 10~\GeV{}.  Furthermore, there must
              be an energy deposit in the hadron calorimeter.

\item{\eemm}: there is no track in the muon  chambers  but  there is one
              track in the central  tracker with a momentum  larger than
              15~\GeV{},  which  points  to an  energy  deposit  in  the
              calorimeters  compatible  with that of a minimum  ionising
              particle.

\item{\ggee}: there is one track in the central  tracker with a momentum
              of less  than  10~\GeV{}  pointing  to an  electromagnetic
              cluster in the BGO of energy less than 10~\GeV{}.

\item{\ggmm}: there is no track in the muon  chambers  but  there is one
              track in the central  tracker with a momentum of less than
              9~\GeV{}  which  points  to  an  energy   deposit  in  the
              calorimeters compatible with the expectation for a minimum
              ionising particle.
\end{itemize}

The  background  candidates  passing  these cuts are  rejected  from the
$\eett$  sample if the  acoplanarity  of the event is less than  5~mrad.
Remaining background from $\eeqq$ and $\ggqq$ is suppressed by requiring
the energy  deposited in the  calorimeters to be between 6 and 25~\GeV{}
whenever there is more than one track in the tagged  hemisphere.  Cosmic
rays  are   suppressed  by  requiring   the  event  time,   measured  by
scintillators,  to be within 5~ns of the beam  crossing  time for events
with a track in the muon chambers or a minimum ionising  particle in the
calorimeters.

A sample of 163\,256 tau decays from  $\eett$  final states is tagged in
one  hemisphere  allowing  to study  the  $\tau$  decay in the  analysis
hemisphere.  The number of tagged hemispheres, $N_\tau$, per data taking
period is listed in  Table~\ref{table:evstat}.  The selection efficiency
is estimated from Monte Carlo simulation to be 76\%.

The  remaining  background  is  determined  from the data for each  year
separately.  It is estimated by  comparing  reference  distributions  in
data and Monte Carlo.

\begin{figure}
    \includegraphics[width=\textwidth]{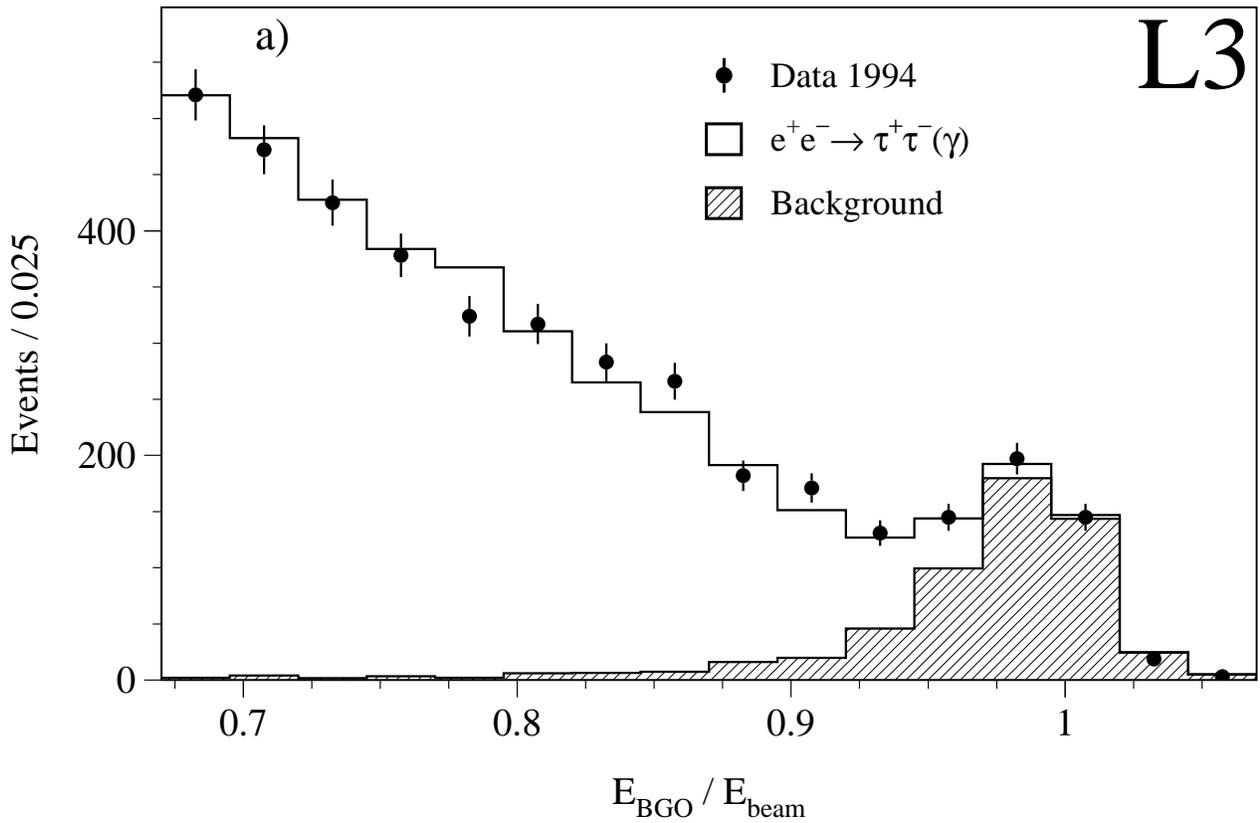}
    \includegraphics[width=\textwidth]{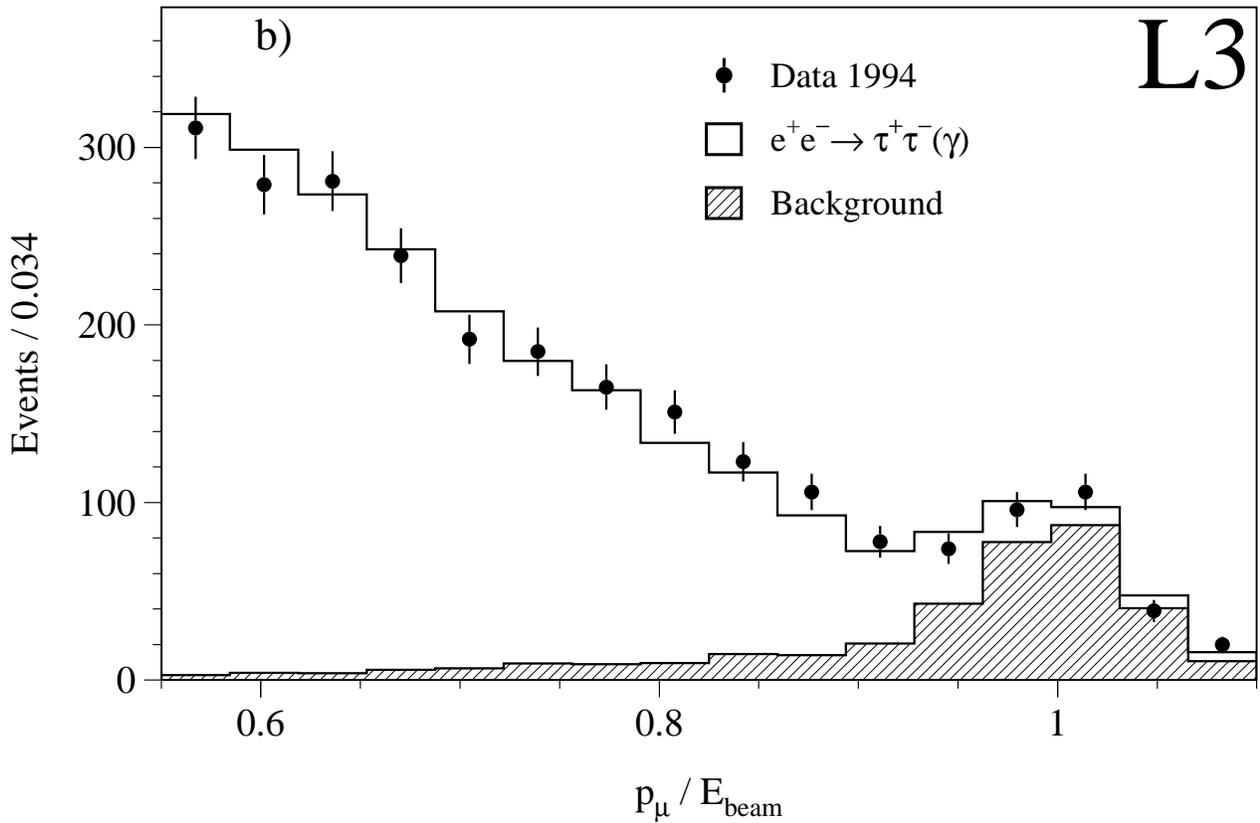}
    \caption[]{\label{fig:ebgo}  Distribution  of a) the  energy  in the
                electromagnetic   calorimeter,  BGO,  and  b)  the  muon
                momentum.  Both  quantities  are  normalised to the beam
                energy,   $\mathrm{E_{beam}}$,   and   measured  in  the
                analysis hemisphere of events tagged as $\eett$.}
\end{figure}

The  fraction  of  $\eeee$  background  is  determined  using the energy
distribution measured in the electromagnetic calorimeter in the analysis
hemisphere, as shown in Figure~\ref{fig:ebgo}a for the 1994 data sample.
The high end of this  spectrum  is  dominated  by $\eeee$  events with a
small  contribution from $\eett$ events.  A fit is performed to the data
distribution,   in  which  the  shapes  of  the   $\eeee$  and   $\eett$
contributions  are taken from Monte Carlo and the  normalisation  of the
$\eeee$ background is a free parameter.

A  similar  procedure  is used for the  other  background  sources.  The
fraction of background  from $\eemm$ events is estimated  using the muon
momentum distribution  measured in the analysis hemisphere, as displayed
for the 1994 data sample in  Figure~\ref{fig:ebgo}b.  The  contamination
from  two-photon   processes  is  estimated   using  the   acollinearity
distribution  and the  fraction  of $\eeqq$  background  by means of the
distribution  of the total  energy  deposited in the  calorimeters.  The
contamination  from cosmic rays is estimated  from data only, using
the  sidebands of the  two-dimensional  distribution  of the distance of
closest approach from the two leading tracks in each hemisphere.

The approximate  contributions  to the $\eett$ sample from the different
background  sources are:  1.2\% from  $\eeee$, 1\% from $\eemm$,  0.25\%
from $\eeqq$, 0.3\% from  two-photon  events and 0.1\% from cosmic rays.
The detailed summary of the estimated  background  fractions is given in
Table~\ref{table:taubkg}.

\section*{Lepton Identification}

The lepton  identification~\cite{Frank}  is  performed  in the  analysis
hemisphere  combining the  informations  from several  subdetectors.  In
particular,  the  electromagnetic   calorimeter  is  essential  for  the
identification  of  electrons,  while the muon  chambers  and the hadron
calorimeter  allow  for  the   identification  of  muons.  Pions  are  a
potential  source of  contamination  both for  electrons and muons.  The
$\rho$ mesons can be  misidentified  as electrons  when the showers from
the charged and neutral pions overlap.

\begin{figure}
    \includegraphics[width=\textwidth]{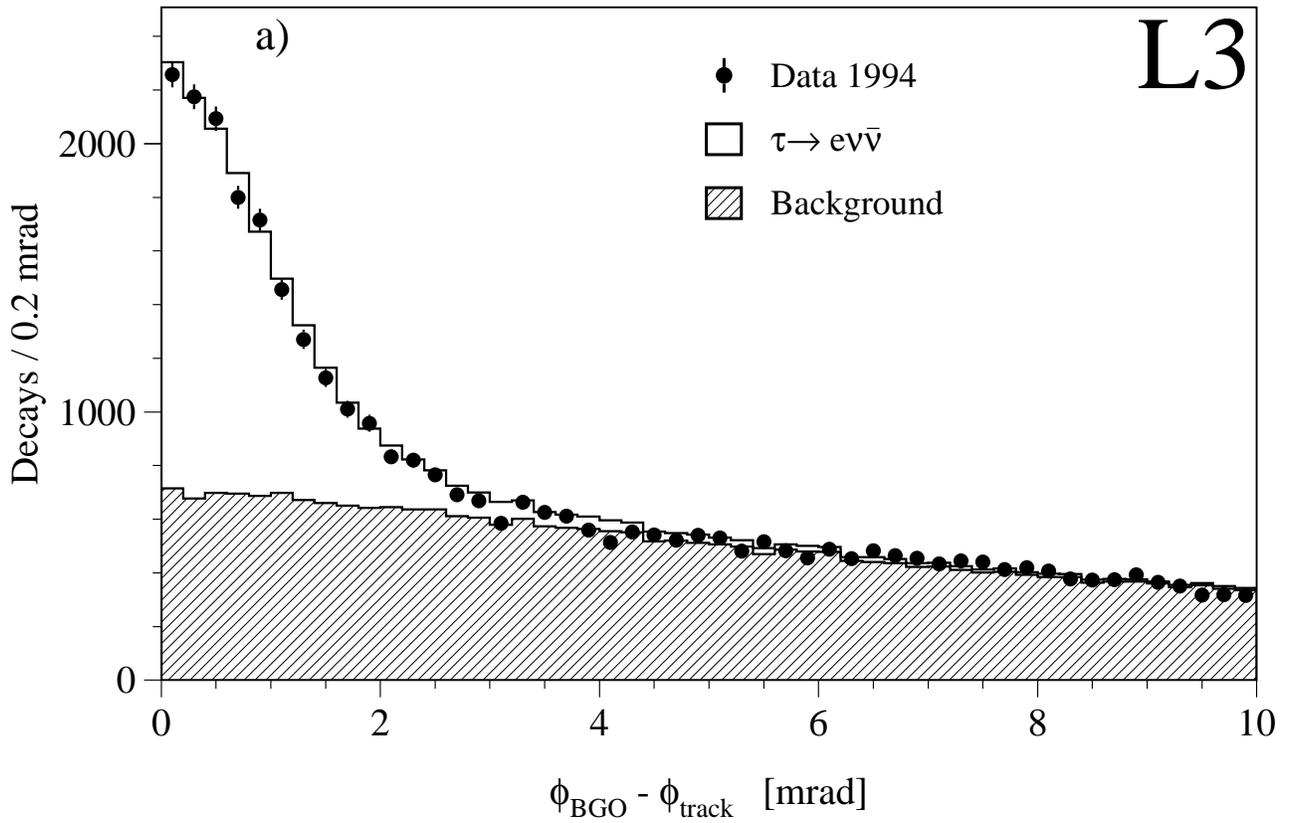}
    \includegraphics[width=\textwidth]{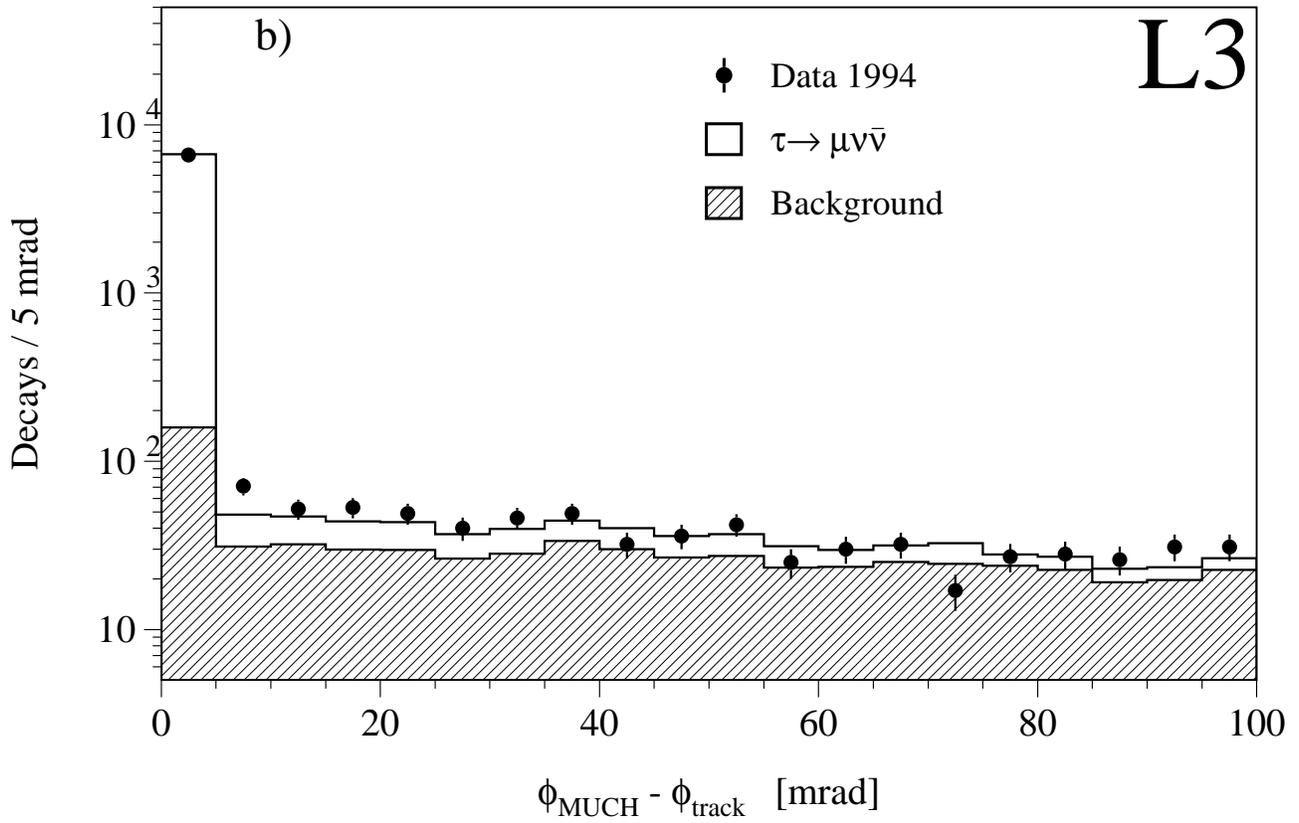}
    \caption[]{\label{fig:phimat}  Distribution of the difference of the
                azimuthal  angles measured by a) the central tracker and
                the  electromagnetic   calorimeter  (BGO)  for  electron
                candidates  and b) the  central  tracker  and  the  muon
                chambers  (MUCH)  for muon  candidates.  The  background
                contributions are shown as the hatched histograms.}
\end{figure}

Electrons are  characterised by a track in the central tracker  pointing
to an energy deposit in the BGO, that must be of electromagnetic  shape.
The  distribution  of the  difference  of the  azimuthal  angles  $\phi$
measured by the central tracker and the  electromagnetic  calorimeter is
shown in  Figure~\ref{fig:phimat}a.  The matching requirement depends on
the electromagnetic energy and varies in $\phi$ between 10 and 3 mrad in
the energy  range from 2 to  45~\GeV.  Similar  cuts are  applied to the
difference in the polar angle.  In addition, the energy  measurement  of
the BGO must be compatible with the track  momentum.  This criterion and
the angular  matching are relaxed for tracks in the anode wire region of
the drift chamber and the  requirements  on the  electromagnetic  shower
shape are tightened.  Energy deposits in the hadron  calorimeter must be
consistent with the tail of an electromagnetic shower.

Muons are  identified as a track in the muon  chambers  matching  with a
track in the central tracker  originating  from the interaction  region.
Furthermore,  the energy deposit in the calorimeters  must correspond to
the  expectations for a minimum ionising  particle.  The distribution of
the difference in the angle $\phi$  measured in the central  tracker and
the muon  chambers  is shown in  Figure~\ref{fig:phimat}b.  Muon  tracks
originating   from  $\tau$  decays  are  well   separated   from  hadron
punch-through.  Muons  reaching  the muon  chambers  lose  energy in the
calorimeters, resulting in a momentum threshold of about 2.5~\GeV{}.

\section*{Efficiencies and Background Estimation}

The efficiencies of the lepton identification estimated from Monte Carlo
are smooth  functions of the lepton  energy and their  average  value is
about 90\% for  electrons  and 75\% for muons.  These  efficiencies  are
corrected using data samples enriched in $\eeee$,  $\eemm$,  $\ggee$ and
$\ggmm$  events,   selected  by  identifying  a  lepton  in  the  tagged
hemisphere.

\begin{figure}
    \includegraphics[width=\textwidth]{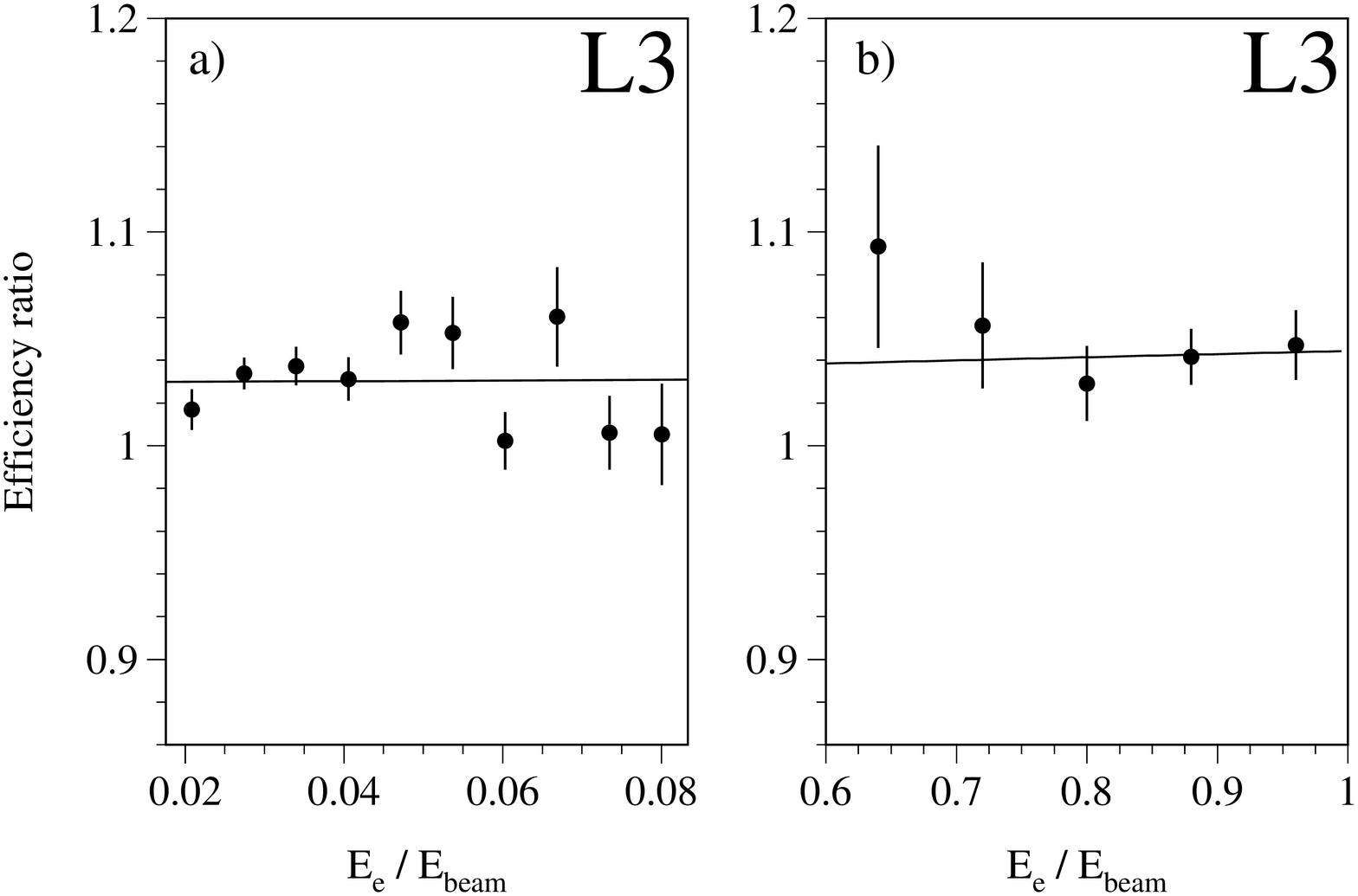}
    \includegraphics[width=\textwidth]{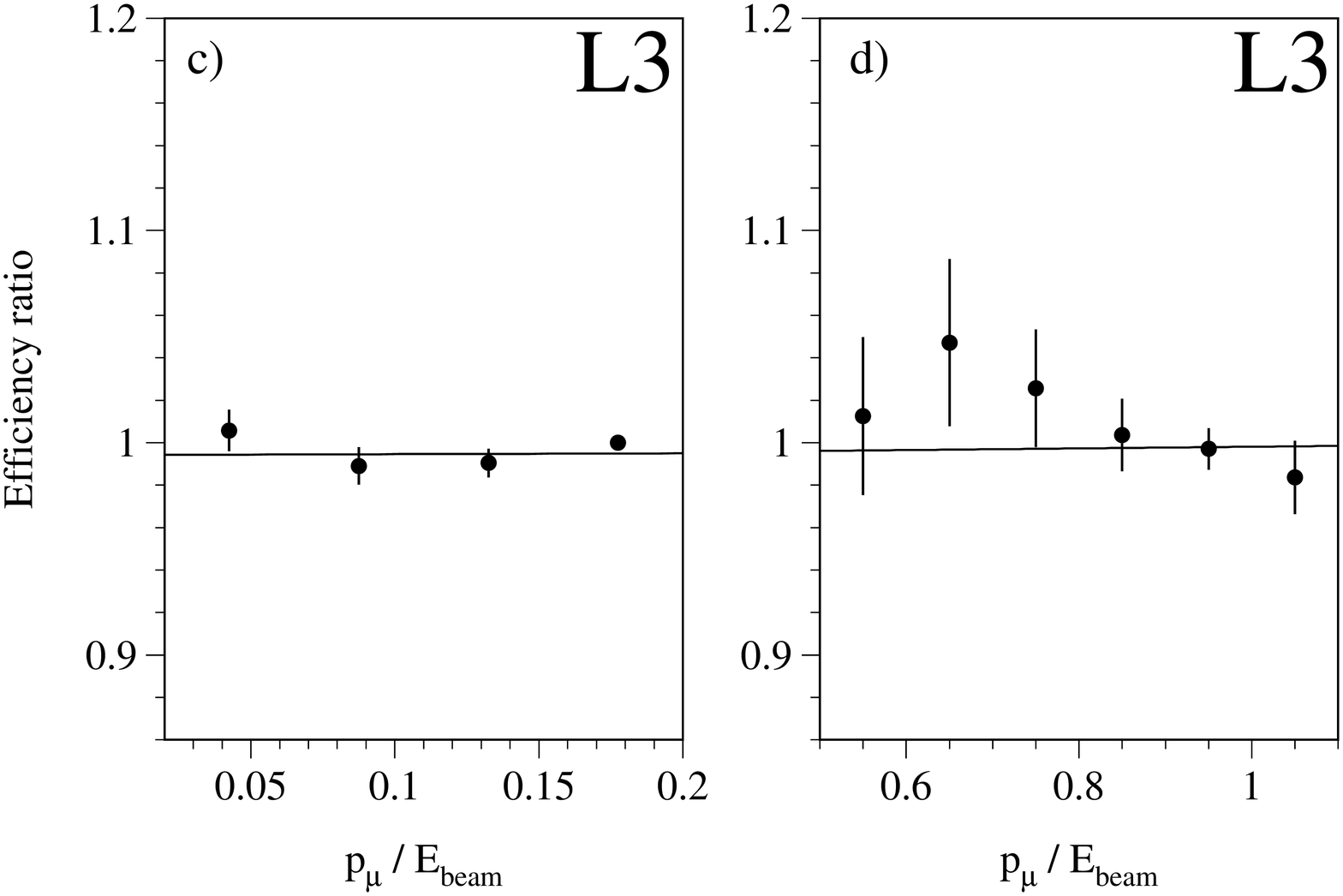}
    \caption[]{\label{fig:dmce}  Ratios  of  the  Monte  Carlo  to  data
                identification efficiencies for electrons and muons as a
                function  of the  normalised  electron  energy  and muon
                momentum.  They are obtained for a)~$\ggee$, b)~$\eeee$,
                c)~$\ggmm$  and d)~$\eemm$  events,  for the 1994  data.
                The straight line is the result of a linear fit over the
                full energy range.}
\end{figure}

The statistics available from $\ggee$ and $\ggmm$ events is large at low
lepton  energy,  while that from $\eeee$ and $\eemm$  events is large at
high  lepton  energy.  Figure~\ref{fig:dmce}  displays  the ratio of the
identification efficiencies obtained from data and Monte Carlo, together
with the result of a linear fit, for  electrons and muons, respectively.
The result of the fit is applied as an energy  dependent scale factor to
the corresponding  identification efficiencies determined from the Monte
Carlo.  These scale factors, obtained for each year separately, are near
unity and almost constant over the full energy range.

The  background  in the  lepton  sample  from  $\eeee$  and  $\eemm$  is
estimated  using the same  procedure as for the $\eett$  selection.  The
fraction of hadronic tau decays  which passes the lepton  identification
is determined from Monte Carlo simulation.

\section*{Systematic Uncertainties}
 
Systematic  uncertainties  result from the preselection,  the cut on the
acollinearity  angle, the  selection of $\eett$  events, the  background
estimations, the lepton  identification  efficiency and the range of the
lepton energy used in the measurement.

The  uncertainties  from the  preselection, the cut on the acollinearity
and the  hemisphere  tagging  criteria  are  estimated  by  varying  the
corresponding  requirements inside reasonable ranges.  The change in the
branching   fraction  is  assigned   as   systematic   uncertainty.  The
background  from $\eeee$,  $\eemm$ and two-photon  processes is obtained
from fits to the data.  The statistical  error of these fits is taken as
the  systematic   uncertainty.  The  systematic   uncertainty  from  the
efficiency  is  obtained  from  the  statistical  errors  of the  energy
dependent  scale  factors.  The Monte Carlo  statistical  uncertainty is
also considered.

These uncertainties are estimated for each year separately~\cite{Frank},
as an  example,  the  uncertainties  for the  1994  data  are  given  in
Table~\ref{table:sys}.  They are  considered as  uncorrelated  and their
combined    values    for   the   full    data   set   are    given   in
Table~\ref{table:sys1}.  This  table  also   presents   the   systematic
uncertainties  fully  correlated  between the data sets of the different
years.  These result from the  background  shapes used in the fit of the
$\eeee$,  $\eemm$ and two-photon  backgrounds and the uncertainty on the
fraction of  misidentified  hadrons.  The  uncertainty  of the one-prong
branching   fraction  of  the  $\tau$  into   hadrons   and  the  $\tau$
polarisation  are also treated as correlated.  Their effect is estimated
by  varying  them  within  their   uncertainties~\cite{pdg20,l3pol}  and
quoting the change of the leptonic branching fraction.

Table~\ref{table:sys2}  lists the  sources of  systematic  uncertainties
correlated  between $\brel$ and $\brmu$.  They are taken into account to
derive $\brlep$ and $\gwm/\gwe$.

\section*{Results}

\begin{figure}
    \includegraphics[width=\textwidth]{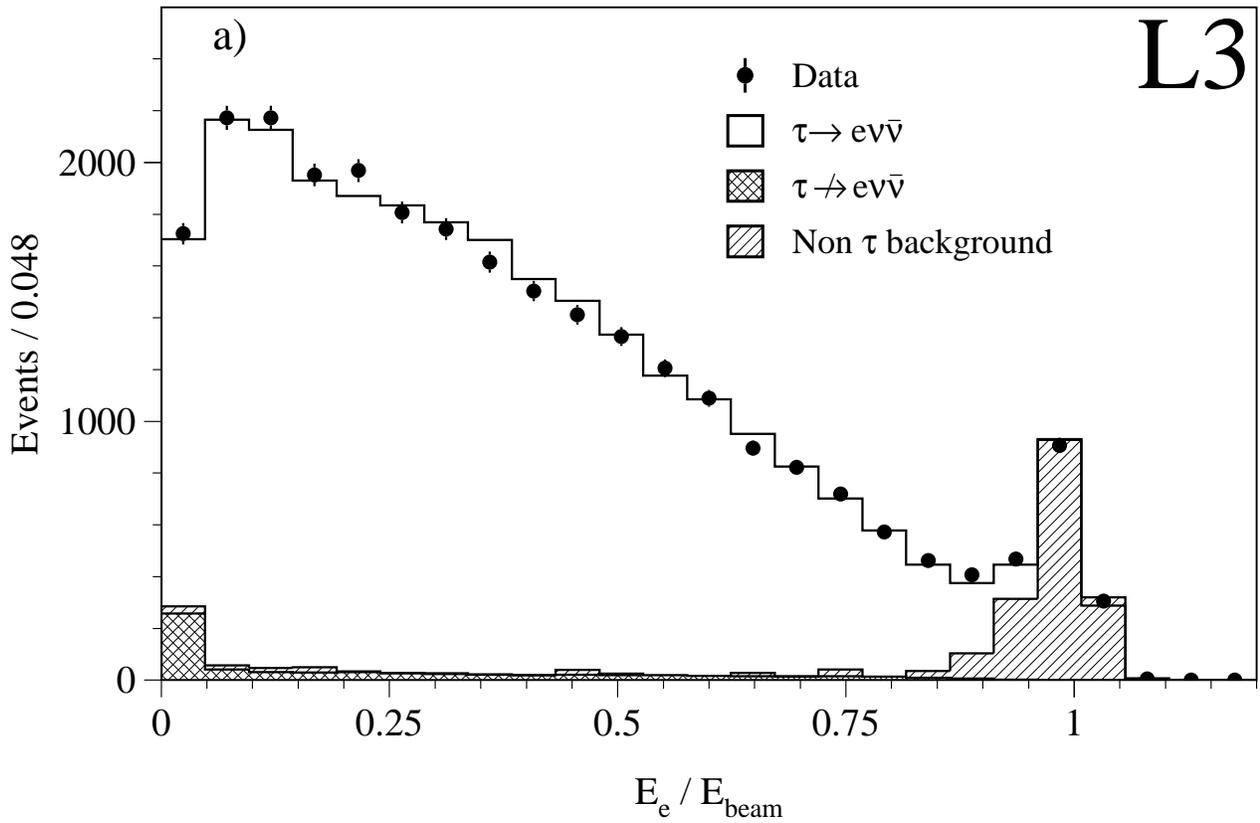}
    \includegraphics[width=\textwidth]{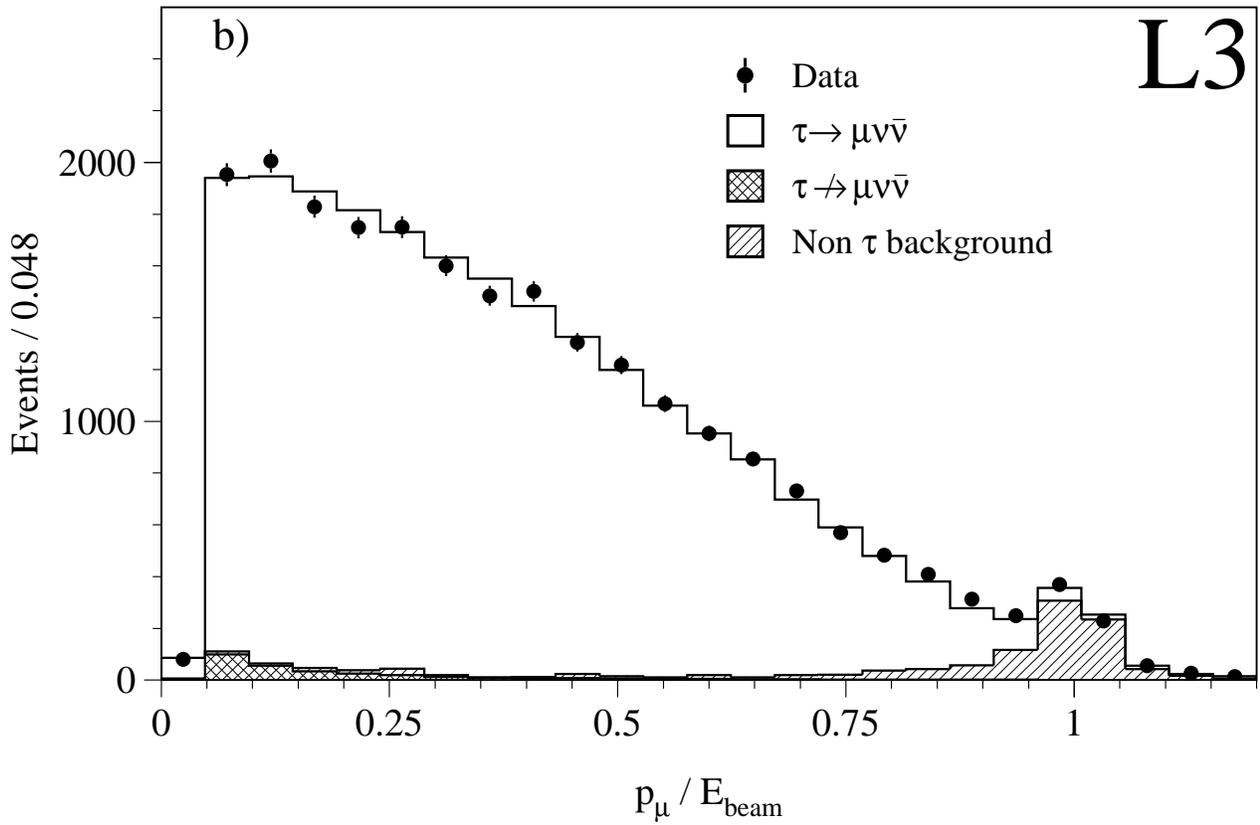}
    \caption[]{\label{fig:espr} The spectra of a) electrons and b) muons
                from  tau   decays   in  the  full   data   sample.  The
                expectations  from the Monte  Carlo  simulation  and the
                background  from  other tau  decays  and other  leptonic
                final states are also shown.}
\end{figure}

Figure~\ref{fig:espr}  displays  the  spectra  of  electrons  and  muons
identified  in the  analysis  hemisphere  in the full data  sample.  The
spectra  obtained  from the  Monte  Carlo  simulations  of  $\tenn$  and
$\tmnn$,  corrected for the  identification  efficiency scale factor and
the  background  sources  are  also  shown.  The  branching  ratios  are
determined  using leptons with  energies  normalised  to the beam energy
that  range  from 0.02 to 0.85 for  electrons  and from 0.05 to 0.92 for
muons.  In  these  ranges  the  efficiencies  are  almost  flat  and the
background  is small.  The number of events  inside  these  ranges,  the
lepton  identification  efficiencies  and the  background  fractions are
given  in  Table~\ref{table:evstat}.  Taking  these  numbers  and  using
Equation~\ref{brform},  the  branching  fractions of the tau lepton into
electron and muon are:
\begin{eqnarray*}
 & \brel  & = (17.806 \pm 0.104 \pm 0.076) \% \quad \mathrm{and} \\
 & \brmu  & = (17.342 \pm 0.110 \pm 0.067) \%,
\end{eqnarray*}
where the first  uncertainty is statistical  and the second  systematic.
These   values   are  in  good   agreement   with  the   current   world
average~\cite{pdg20}.  The results are used to test lepton  universality
for  the  charged  weak  currents.  The  ratio  of the  charged  current
coupling constants of the muon and the electron, is obtained as:
$$\gwm/\gwe = 1.0007  \pm  0.0043  \pm  0.0027,$$
supporting the lepton universality  hypothesis.  Assuming  electron-muon
universality, the branching fraction of the tau into leptons is:
$$\brlep = (17.818  \pm  0.077  \pm  0.053)~\%.$$
Together  with  our  measurement  of the tau  lifetime~\cite{lifet}  the
Fermi constant in tau lepton decays is determined as:
$$\gf = (1.1616 \pm 0.0058)\times 10^{-5} \GeV^{-2}.$$
Furthermore,  from  the  branching  fraction  of the tau  into  leptons,
$\rtau$ is obtained using Equation~\ref{eq:rtau} as:
$$\rtau=3.640~\pm~0.030.$$
From Equation~\ref{eq:als}, the value of the strong coupling constant at
the tau mass is:
$$\als(\Mtau^2)=0.322~\pm~0.009~\mbox{(exp.)}~\pm~0.015~\mbox{(theory)}.$$
The  first  error  is  due  to the  uncertainties  of the  tau  leptonic
branching fraction and the CKM matrix elements~\cite{pdg20}.  The second
error is the  quadratic  sum of the  uncertainties  resulting  from  the
renormalisation  scale, the fourth order term in $\als$, the electroweak
corrections $\sew$ and the the  non-perturbative  correction  $\dnpqcd$.
The  dominant  contribution  to the error is the  renormalisation  scale
uncertainty,  which  is  estimated  following  Reference~\citen{pich92}.
Other theoretical  uncertainties  discussed in  Reference~\citen{alta95}
are not considered.  The value of $\als(\Mtau^2)$ is extrapolated to the
$\Zo$ mass scale using the  renormalisation  group  equation~\cite{rge1}
with the four loop calculation of the QCD  $\beta$-functions~\cite{rge}.
The result,
$$\als(m_{\Zo}^2) = 0.120~\pm~0.002,$$
is in good  agreement  with the value  obtained  by L3 from the study of
hadronic  events at the Z  peak~\cite{alpha_l3}  and the  current  world
average~\cite{pdg20}.

\section*{Acknowledgements}

We  thank  A.  Kataev  for  discussions  about  the  estimation  of  the
theoretical uncertainty of $\rtau$.  We wish to express our gratitude to
the CERN accelerator  divisions for the excellent performance of the LEP
machine.  We  acknowledge  the   contributions   of  the  engineers  and
technicians who have participated in the construction and maintenance of
this experiment.

\newpage
\typeout{   }     
\typeout{Using author list for paper 233 only}
\typeout{$Modified: Jan 25 2000 by smele $}
\typeout{!!!!  This should only be used with document option a4p!!!!}
\typeout{   }
%
%
%
%
%
%

\newcount\tutecount  \tutecount=0
\def\tutenum#1{\global\advance\tutecount by 1 \xdef#1{\the\tutecount}}
\def\tute#1{$^{#1}$}
\tutenum\aachen            
\tutenum\nikhef            
\tutenum\mich              
\tutenum\lapp              
\tutenum\basel             
\tutenum\lsu               
\tutenum\beijing           
\tutenum\berlin            
\tutenum\bologna           
\tutenum\tata              
\tutenum\ne                
\tutenum\bucharest         
\tutenum\budapest          
\tutenum\mit               
\tutenum\debrecen          
\tutenum\florence          
\tutenum\cern              
\tutenum\wl                
\tutenum\geneva            
\tutenum\hefei             
\tutenum\lausanne          
\tutenum\lecce             
\tutenum\lyon              
\tutenum\madrid            
\tutenum\milan             
\tutenum\moscow            
\tutenum\naples            
\tutenum\cyprus            
\tutenum\nymegen           
\tutenum\caltech           
\tutenum\perugia           
\tutenum\peters            
\tutenum\cmu               
\tutenum\potenza           
\tutenum\prince            
\tutenum\riverside         
\tutenum\rome              
\tutenum\salerno           
\tutenum\ucsd              
\tutenum\sofia             
\tutenum\korea             
\tutenum\alabama           
\tutenum\utrecht           
\tutenum\purdue            
\tutenum\psinst            
\tutenum\zeuthen           
\tutenum\eth               
\tutenum\hamburg           
\tutenum\taiwan            
\tutenum\tsinghua          

{
\parskip=0pt
\noindent
{\bf The L3 Collaboration:}
\ifx\selectfont\undefined
 \baselineskip=10.8pt
 \baselineskip\baselinestretch\baselineskip
 \normalbaselineskip\baselineskip
 \ixpt
\else
 \fontsize{9}{10.8pt}\selectfont
\fi
\medskip
\tolerance=10000
\hbadness=5000
\raggedright
\hsize=162truemm\hoffset=0mm
\def\r{\rlap,}
\noindent

M.Acciarri\r\tute\milan\
P.Achard\r\tute\geneva\ 
O.Adriani\r\tute{\florence}\ 
M.Aguilar-Benitez\r\tute\madrid\ 
J.Alcaraz\r\tute\madrid\ 
G.Alemanni\r\tute\lausanne\
J.Allaby\r\tute\cern\
A.Aloisio\r\tute\naples\ 
M.G.Alviggi\r\tute\naples\
G.Ambrosi\r\tute\geneva\
% ...
\newpage
\begin{list}{A}{\itemsep=0pt plus 0pt minus 0pt\parsep=0pt plus 0pt minus 0pt
                \topsep=0pt plus 0pt minus 0pt}
\item[\aachen]
 I. Physikalisches Institut, RWTH, D-52056 Aachen, FRG$^{\S}$\\
 III. Physikalisches Institut, RWTH, D-52056 Aachen, FRG$^{\S}$
\item[\nikhef] National Institute for High Energy Physics, NIKHEF, 
     and University of Amsterdam, NL-1009 DB Amsterdam, The Netherlands
\item[\mich] University of Michigan, Ann Arbor, MI 48109, USA
\item[\lapp] Laboratoire d'Annecy-le-Vieux de Physique des Particules, 
     LAPP,IN2P3-CNRS, BP 110, F-74941 Annecy-le-Vieux CEDEX, France
\item[\basel] Institute of Physics, University of Basel, CH-4056 Basel,
     Switzerland
\item[\lsu] Louisiana State University, Baton Rouge, LA 70803, USA
\item[\beijing] Institute of High Energy Physics, IHEP, 
  100039 Beijing, China$^{\triangle}$ 
\item[\berlin] Humboldt University, D-10099 Berlin, FRG$^{\S}$
\item[\bologna] University of Bologna and INFN-Sezione di Bologna, 
     I-40126 Bologna, Italy
\item[\tata] Tata Institute of Fundamental Research, Bombay 400 005, India
\item[\ne] Northeastern University, Boston, MA 02115, USA
\item[\bucharest] Institute of Atomic Physics and University of Bucharest,
     R-76900 Bucharest, Romania
\item[\budapest] Central Research Institute for Physics of the 
     Hungarian Academy of Sciences, H-1525 Budapest 114, Hungary$^{\ddag}$
\item[\mit] Massachusetts Institute of Technology, Cambridge, MA 02139, USA
\item[\debrecen] KLTE-ATOMKI, H-4010 Debrecen, Hungary$^\P$
\item[\florence] INFN Sezione di Firenze and University of Florence, 
     I-50125 Florence, Italy
\item[\cern] European Laboratory for Particle Physics, CERN, 
     CH-1211 Geneva 23, Switzerland
\item[\wl] World Laboratory, FBLJA  Project, CH-1211 Geneva 23, Switzerland
\item[\geneva] University of Geneva, CH-1211 Geneva 4, Switzerland
\item[\hefei] Chinese University of Science and Technology, USTC,
      Hefei, Anhui 230 029, China$^{\triangle}$
\item[\lausanne] University of Lausanne, CH-1015 Lausanne, Switzerland
\item[\lecce] INFN-Sezione di Lecce and Universit\`a Degli Studi di Lecce,
     I-73100 Lecce, Italy
\item[\lyon] Institut de Physique Nucl\'eaire de Lyon, 
     IN2P3-CNRS,Universit\'e Claude Bernard, 
     F-69622 Villeurbanne, France
\item[\madrid] Centro de Investigaciones Energ{\'e}ticas, 
     Medioambientales y Tecnolog{\'\i}cas, CIEMAT, E-28040 Madrid,
     Spain${\flat}$ 
\item[\milan] INFN-Sezione di Milano, I-20133 Milan, Italy
\item[\moscow] Institute of Theoretical and Experimental Physics, ITEP, 
     Moscow, Russia
\item[\naples] INFN-Sezione di Napoli and University of Naples, 
     I-80125 Naples, Italy
\item[\cyprus] Department of Natural Sciences, University of Cyprus,
     Nicosia, Cyprus
\item[\nymegen] University of Nijmegen and NIKHEF, 
     NL-6525 ED Nijmegen, The Netherlands
\item[\caltech] California Institute of Technology, Pasadena, CA 91125, USA
\item[\perugia] INFN-Sezione di Perugia and Universit\`a Degli 
     Studi di Perugia, I-06100 Perugia, Italy   
\item[\peters] Nuclear Physics Institute, St. Petersburg, Russia
\item[\cmu] Carnegie Mellon University, Pittsburgh, PA 15213, USA
\item[\potenza] INFN-Sezione di Napoli and University of Potenza, 
     I-85100 Potenza, Italy
\item[\prince] Princeton University, Princeton, NJ 08544, USA
\item[\riverside] University of Californa, Riverside, CA 92521, USA
\item[\rome] INFN-Sezione di Roma and University of Rome, ``La Sapienza",
     I-00185 Rome, Italy
\item[\salerno] University and INFN, Salerno, I-84100 Salerno, Italy
\item[\ucsd] University of California, San Diego, CA 92093, USA
\item[\sofia] Bulgarian Academy of Sciences, Central Lab.~of 
     Mechatronics and Instrumentation, BU-1113 Sofia, Bulgaria
\item[\korea]  Laboratory of High Energy Physics, 
     Kyungpook National University, 702-701 Taegu, Republic of Korea
\item[\alabama] University of Alabama, Tuscaloosa, AL 35486, USA
\item[\utrecht] Utrecht University and NIKHEF, NL-3584 CB Utrecht, 
     The Netherlands
\item[\purdue] Purdue University, West Lafayette, IN 47907, USA
\item[\psinst] Paul Scherrer Institut, PSI, CH-5232 Villigen, Switzerland
\item[\zeuthen] DESY, D-15738 Zeuthen, 
     FRG
\item[\eth] Eidgen\"ossische Technische Hochschule, ETH Z\"urich,
     CH-8093 Z\"urich, Switzerland
\item[\hamburg] University of Hamburg, D-22761 Hamburg, FRG
\item[\taiwan] National Central University, Chung-Li, Taiwan, China
\item[\tsinghua] Department of Physics, National Tsing Hua University,
      Taiwan, China
\item[\S]  Supported by the German Bundesministerium 
        f\"ur Bildung, Wissenschaft, Forschung und Technologie
\item[\ddag] Supported by the Hungarian OTKA fund under contract
numbers T019181, F023259 and T024011.
\item[\P] Also supported by the Hungarian OTKA fund under contract
  numbers T22238 and T026178.
\item[$\flat$] Supported also by the Comisi\'on Interministerial de Ciencia y 
        Tecnolog{\'\i}a.
\item[$\sharp$] Also supported by CONICET and Universidad Nacional de La Plata,
        CC 67, 1900 La Plata, Argentina.
\item[$\diamondsuit$] Also supported by Panjab University, Chandigarh-160014, 
        India.
\item[$\natural$] Also supported by Deutscher akademischer 
Austauschdienst.
\item[$\triangle$] Supported by the National Natural Science
  Foundation of China.
\end{list}
}
\vfill


\newpage

\begin{table}[htbp]
\begin{center}
\begin{tabular}{|c|r@{$\,\pm\,$}l|r@{$\,\pm\,$} l|r@{$\,\pm\,$} l|r@{$\,\pm\,$}l|}
\cline{2-9}
\multicolumn{1}{c|}{}          & \multicolumn{2}{|c|}{1991+92} & \multicolumn{2}{|c|}{1993}    & \multicolumn{2}{|c|}{1994}    & \multicolumn{2}{|c|}{1995}    \\ \hline
\multicolumn{1}{|c|}{$N_\tau$} & \multicolumn{2}{|c|}{42\,086} & \multicolumn{2}{|c|}{29\,620} & \multicolumn{2}{|c|}{65\,392} & \multicolumn{2}{|c|}{26\,158} \\ \hline
\multicolumn{1}{|c|}{$N_e$ (\tenn)} & \multicolumn{2}{|c|}{6\,519}  & \multicolumn{2}{|c|}{4\,525}  & \multicolumn{2}{|c|}{9\,754}  & \multicolumn{2}{|c|}{3\,878}  \\ \hline
\multicolumn{1}{|c|}{\rule{0pt}{12pt}$\varepsilon^{ID}_{e}$ ($\%$)} & 85.94 & 0.26 & 83.62 & 0.29 & 83.86 & 0.17 & 82.75 & 0.26 \\
\multicolumn{1}{|c|}{$f_e^{non-\tau}$ ($\%$)}                       &  1.43 & 0.23 &  2.42 & 0.38 &  1.32 & 0.19 &  2.11 & 0.34 \\
\multicolumn{1}{|c|}{$f_e^{\tau \nrightarrow e}$ ($\%$)}            &  2.04 & 0.12 &  1.74 & 0.11 &  1.70 & 0.10 &  1.76 & 0.10 \\ \hline
\multicolumn{1}{|c|}{$N_\mu$ (\tmnn)}& \multicolumn{2}{|c|}{5\,682} & \multicolumn{2}{|c|}{3\,984} & \multicolumn{2}{|c|}{8\,554} & \multicolumn{2}{|c|}{3\,289} \\ \hline
\multicolumn{1}{|c|}{\rule{0pt}{12pt}$\varepsilon^{ID}_{\mu}$ ($\%$)} & 76.22 & 0.23 & 77.65 & 0.24 & 75.10 & 0.17 & 75.54 & 0.24 \\
\multicolumn{1}{|c|}{$f_\mu^{non-\tau}$ ($\%$)}                       &  1.29 & 0.16 &  1.89 & 0.23 &  1.11 & 0.09 &  1.60 & 0.15 \\
\multicolumn{1}{|c|}{$f_\mu^{\tau \nrightarrow \mu}$ ($\%$)}          &  1.41 & 0.09 &  1.38 & 0.09 &  1.44 & 0.09 &  1.17 & 0.08 \\ \hline
\end{tabular}
\caption{Number  of selected tau decays, number of identified  electrons
         and muons,  lepton  identification  efficiencies,  fraction  of
         background from non-$\tau$ final states and other $\tau$ decays
         for the different data taking periods.}
\label{table:evstat} 
\end{center}
\end{table}

\begin{table}[htp]
\begin{center}
\begin{tabular}{|l|l@{$\,\pm\,$}l| l@{$\,\pm\,$} l| l@{$\,\pm\,$} l| l @{$\,\pm\,$}l|} \cline{2-9}
 \multicolumn{1}{c|}{} & \multicolumn{8}{c|}{Background in \%}  \\ \hline
 \multicolumn{1}{|c|}{Channel}    & \multicolumn{2}{c|}{1991+92} & \multicolumn{2}{c|}{1993} &
\multicolumn{2}{c|}{1994} & \multicolumn{2}{c|}{1995} \\ 
\hline
\eemm  & 0.96 &  0.08 & 0.97 &  0.09 & 0.94 &  0.05 & 1.11 &  0.08 \\
\eeee  & 1.19 &  0.14 & 1.67 &  0.19 & 1.16 &  0.07 & 1.33 &  0.11 \\ 
\ggmm  & 0.03 &  0.01 & 0.12 &  0.02 & 0.03 &  0.01 & 0.14 &  0.02 \\
\ggee  & 0.15 &  0.03 & 0.27 &  0.04 & 0.15 &  0.02 & 0.29 &  0.05 \\
\ggqq  & 0.04 &  0.02 & 0.06 &  0.03 & 0.07 &  0.02 & 0.06 &  0.03 \\
Cosmic rays
       & 0.04 &  0.01 & 0.11 &  0.02 & 0.10 &  0.01 & 0.17 &  0.03 \\ 
\eeqq  & 0.22 &  0.05 & 0.23 &  0.06 & 0.24 &  0.05 & 0.14 &  0.03 \\
\hline 

Total & 2.64 & 0.17 & 3.42 &  0.23 & 2.70 & 0.10 & 3.24 & 0.15 \\ 
\hline 
\end{tabular}
\end{center}
\caption{Summary of the background fractions in the $\eett$~sample.}
\label{table:taubkg} 
\end{table}

\begin{table}[htp]
\begin{center}
\begin{tabular}{|l|c|c|}
\hline 
Source   &  \brel & \brmu  \\
\hline
Preselection and acollinearity angle & 0.040   & 0.031  \\
Selection of \eett                   & 0.026   & 0.029  \\
Uncorrelated background              & 0.031   & 0.021  \\
Lepton identification scale factor   & 0.041   & 0.048  \\
Lepton energy range                  & 0.054   & 0.028  \\
Monte Carlo statistics               & 0.053   & 0.057  \\
\hline
Total uncertainty                    & 0.103   & 0.093 \\                        \hline 
\end{tabular}
\caption{Uncorrelated  systematic uncertainties for the 1994 data sample
         and their combination.}
\label{table:sys} 
\end{center}
\end{table}

\begin{table}[htp]
\begin{center}
\begin{tabular}{|l|c|c|}
\hline 
Source   &  \brel & \brmu  \\
\hline
Correlated background                & 0.036   & 0.025  \\
Hadronic 1-prong branching fraction  & 0.021   & 0.018  \\
Polarisation                         & 0.003   & 0.003  \\
\hline
Total   correlated uncertainty       & 0.042   & 0.031  \\
Total uncorrelated uncertainty       & 0.063   & 0.059  \\
\hline
Total uncertainty                    & 0.076   & 0.067  \\                       \hline 
\end{tabular}
\caption{Correlated  and uncorrelated  systematic  uncertainties for the
         full data sample.}
\label{table:sys1} 
\end{center}
\end{table}

\begin{table}[htp]
\begin{center}
\begin{tabular}{|l|c|}
\hline 
Source   &  \brlep \\
\hline
Background in the $\tau$ sample       & 0.013   \\
Hadronic 1-prong branching fraction   & 0.013   \\
Polarisation                          & 0.003   \\
Acollinearity angle                   & 0.012   \\
Selection                             & 0.004   \\
\hline
Total uncertainty                     & 0.023    \\                     
\hline 
\end{tabular}
\caption{Systematic   uncertainties   on   $\brlep$   treated  as  fully
         correlated for the electron and muon channels.}
\label{table:sys2} 
\end{center}
\end{table}

\end{document}